\documentclass[final,1p,times]{elsarticle} 
\usepackage{graphicx} 
\usepackage{amssymb} 
\usepackage{amsthm} 
\usepackage{lineno} 
\pdfoutput=1

\journal{Nuclear Physics A} 
\begin{document} 

\begin{frontmatter} 


\title{Open Heavy Flavor Measurements at PHENIX}

\author{Alan Dion$^{a}$ for the PHENIX collaboration}

\address{Iowa State University, Department of Physics and Astronomy}

\begin{abstract} 
We report on two new measurements by the PHENIX collaboration regarding open heavy flavor.  The first is an additional contribution to the background in the measurement of the yield of electrons at mid-rapidity from open heavy flavor decays.  We found that the $J/\psi$ can contribute substantially to the yield at high $p_T$.  PHENIX has also measured the spectrum of muons at forward rapidity from open heavy flavor decays in Cu+Cu collisions at $\sqrt{s_{NN}}$ = 200 GeV.

\end{abstract} 

\end{frontmatter} 


\section{New Background Contributions to Electrons from Open Heavy Flavor Decays at Mid-Rapidity}

PHENIX has measured the non-photonic electron spectrum in $p$+$p$ \cite{e_pp} and Au+Au \cite{e_aa} at $\sqrt{s_{NN}}$ = 200 GeV in the pseudorapidity range $|\eta|<0.35$.  The non-photonic electrons have been interpreted as coming almost entirely from decays of open heavy flavor hadrons.  PHENIX has now calculated that approximately 16\% of the electron signal at high $p_T$ actually comes from $J/\psi$ decays.

\begin{figure}[!h]
\centering
\includegraphics[width=0.45\textwidth]{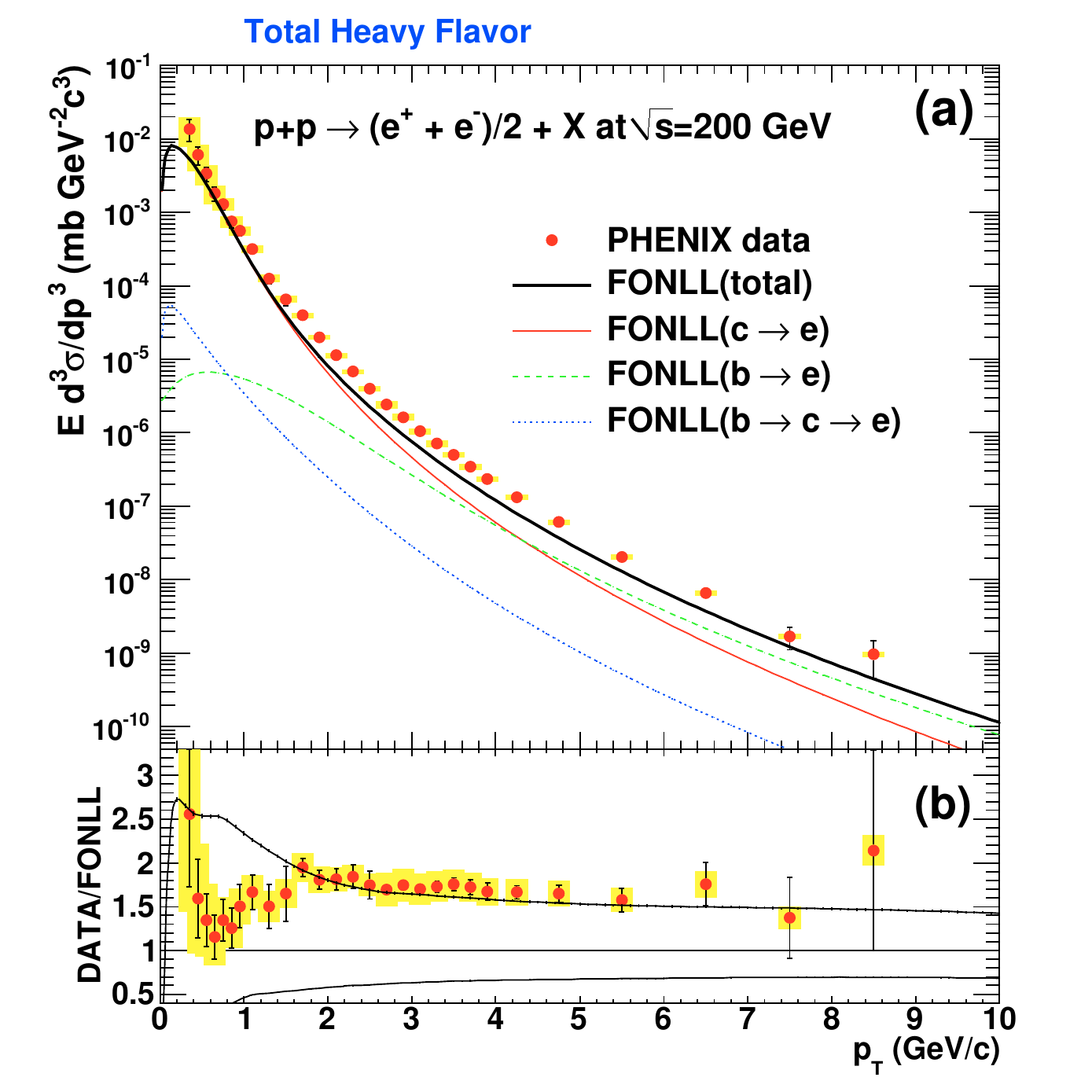}
\includegraphics[width=0.45\textwidth]{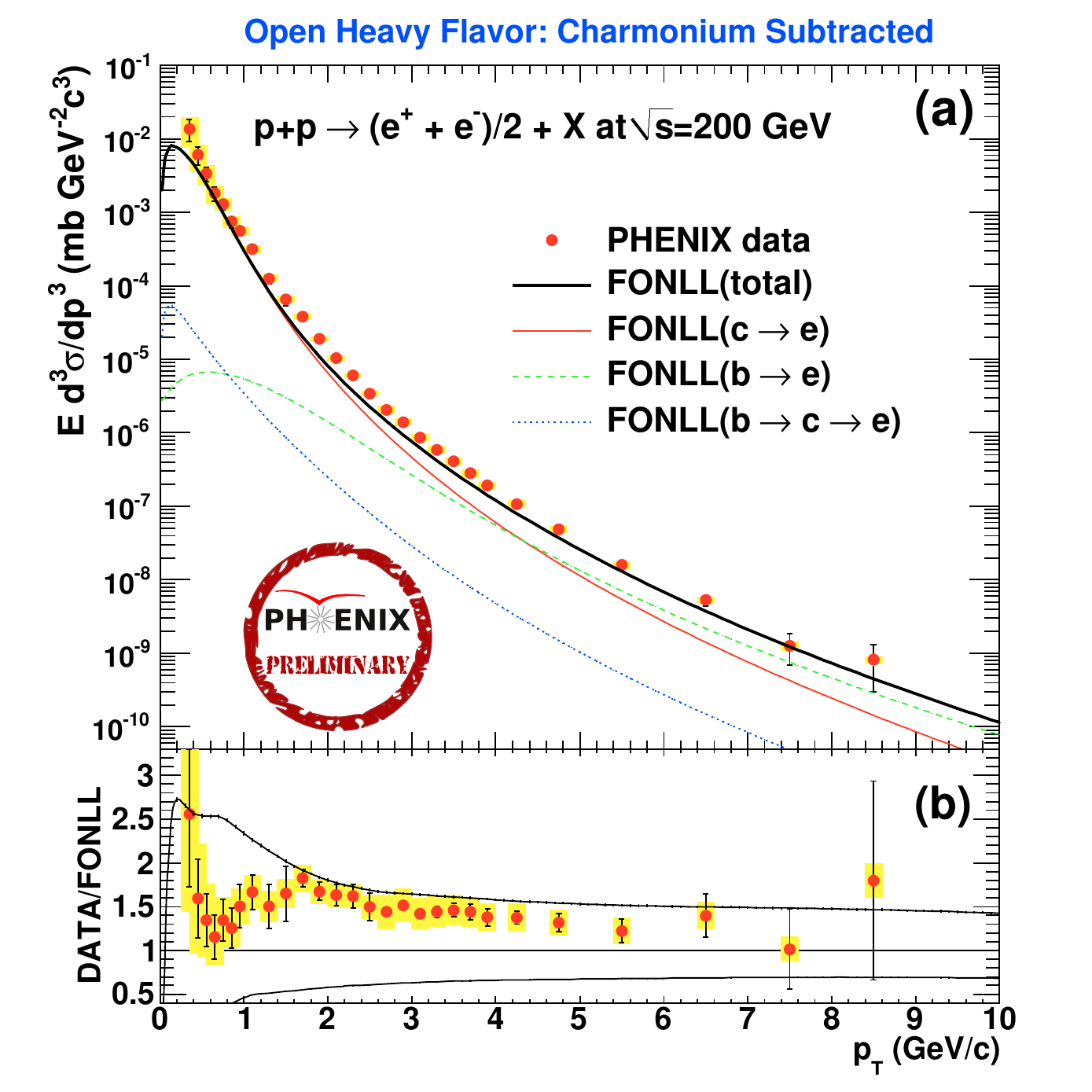}

\caption{left: Non-photonic electron yield compared to an FONLL calculation
\cite{fonll}  right: The non-photonic electron yield after subtraction of the
$J/\psi$, $\Upsilon$, and Drell-Yan contributions.}
\label{fig:ppyield}
\end{figure}

The $J/\psi$ $p_T$ spectrum from $p$+$p$ collisions has been measured by PHENIX out to 9 GeV/$c$ \cite{cesar}.  The data from this measurement is used as an input to a Monte Carlo event generator to determine the yield of electrons from the decay of $J/\psi$ particles.  To determine the appropriate input to the event generator, two functional fits to the $J/\psi$ data are performed.  The first uses the so-called Kaplan function of the form $p_0 \left( 1 + \left( p_T/p_1 \right)^2 \right)^{-n}$, and the second uses the $m_T$ scaling function used in \cite{mt}.  The values of the electron spectrum from the event generator using the two different functional forms are averaged.  The systematic errors on the $J/\psi$ measurement are propagated to errors on the electron spectrum.
\begin{figure}[!h]
\centering
\includegraphics[width=0.43\textwidth]{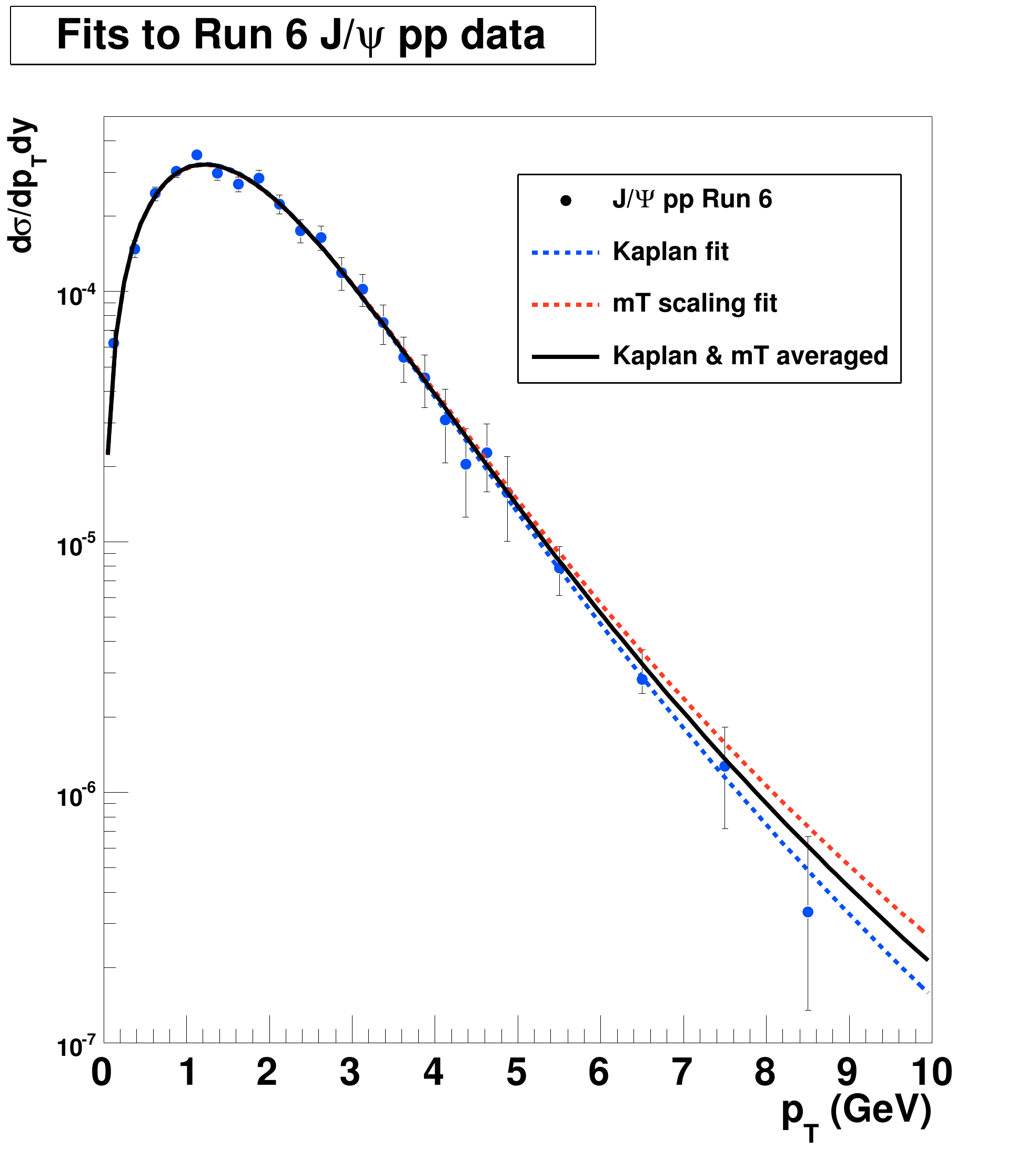}
\includegraphics[width=0.5\textwidth]{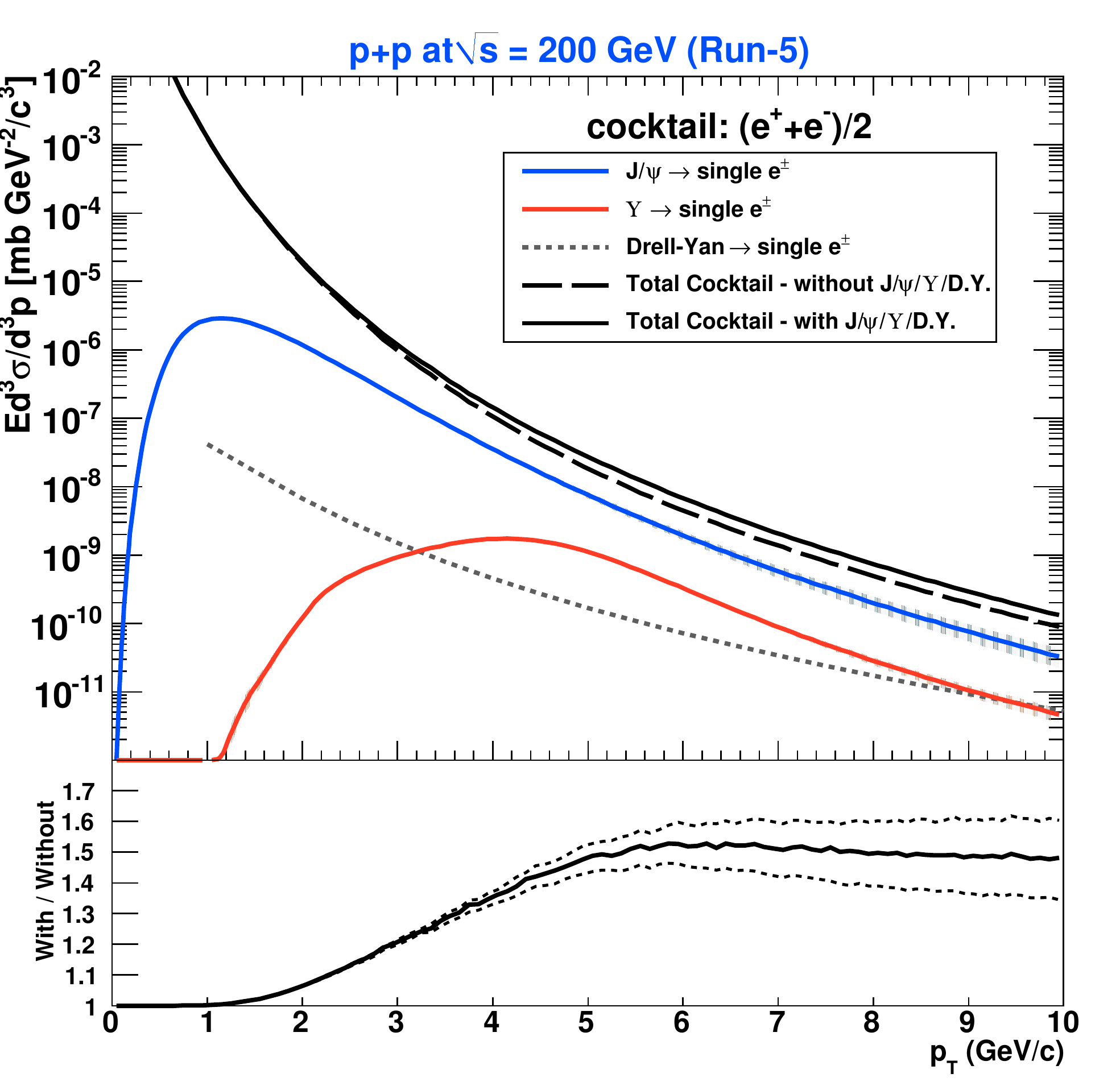}

\caption{left: Fits of two functional forms to the $p$+$p$ $J/\psi$ data. 
right: Comparisons of the electron background "cocktail" before and after the
addition of the new components.}
\label{fig:ppfits}
\end{figure}
Figure \ref{fig:ppfits} shows the fits to the $J/\psi$ data and the change in
the electron background.  Although the background at the highest $p_T$ is
increased by 50\%, the signal to background is well above 1 in that $p_T$ range
\cite{e_pp}, and the change in the heavy flavor signal is less than 20\%, as can
be seen from Figure \ref{fig:ppyield}.
Estimations for the contribution from Upsilon and Drell-Yan were also
considered, but were found to be quite small.

\begin{figure}[!h]
\centering
\includegraphics[width=0.55\textwidth]{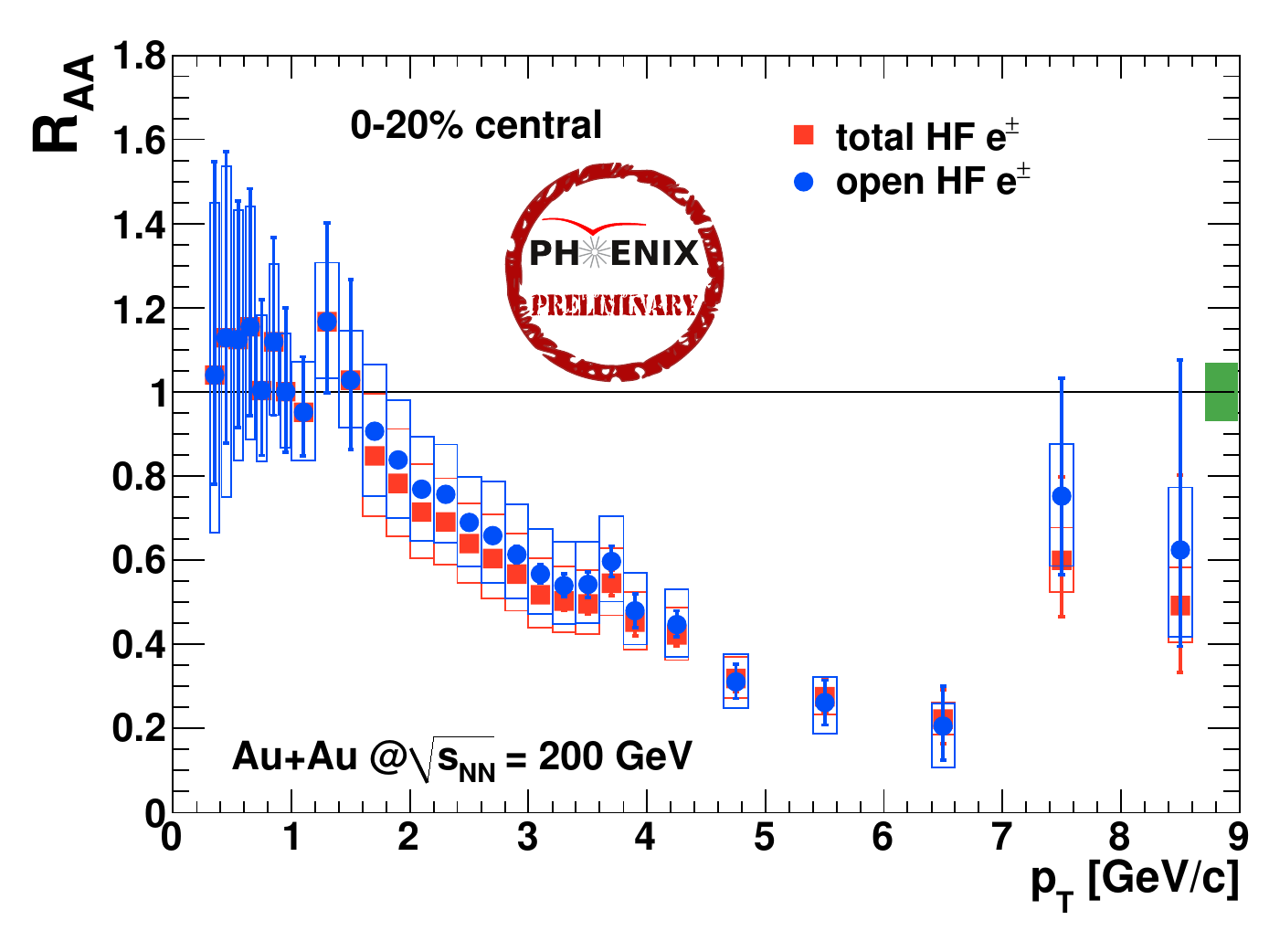}

\caption{$R_{AA}$ of non-photonic electrons before ("total") and after ("open")
subtraction of the $J/\psi$ contributions. }
\label{fig:eraa}
\end{figure}

In the case of Au+Au collisions, the $J/\psi$ is only measured in PHENIX out to
$p_T$ of about 5 GeV/$c$.  In order to gain an estimation of the $J/\psi$
contribution to the electron spectrum in Au+Au, we made two conservative
estimations of the upper and lower bound of the $J/\psi$ yield at high $p_T$
using the measured $R_{AA}$ \cite{jraa} in the 0-20\% centrality range.  For the
lower bound on the $J/\psi$ yield, we assume that the $R_{AA}$ stays constant at
about 0.34 out to 10 GeV/$c$, and for the upper bound we assumed that it
increases linearly from 0.34 to 1.0 between 5 GeV/$c$ and 10 GeV/$c$.  The
resulting $R_{AA}$ for single electrons for 0-20\% centrality with and without
the $J/\psi$ contributions can be seen in Figure \ref{fig:eraa}.
As both the $p$+$p$ and Au+Au single electron yields from heavy flavor decays
decreased as a result of subtracting the $J/\psi$ contribution, the resulting
change in $R_{AA}$ is small.  Note that PHENIX previously highlighted the
$R_{AA}$ in 0-10\% centrality \cite{e_aa}, but here we show the 0-20\%
centrality range as that is how the $J/\psi$ was measured.

\section{Single Muon Measurement in Cu+Cu collisions}

PHENIX has previously measured the yield of prompt single muons at forward
rapidity in $p$+$p$ collisions \cite{donny, muon}.  These prompt muons are
attributed to semileptonic decays of open heavy flavor hadrons.  PHENIX has now
measured the yield of single muons from Cu+Cu collisions at $\sqrt{s_{NN}}$ =
200 GeV in the pseudorapidity range $1.4<|\eta|<1.9$.  The measurement
methodology used for Cu+Cu collisions is very similar to what was used for the
$p$+$p$ collisions.  Perhaps the largest change for the Cu+Cu is the calculation
of multiplicity-dependent efficiency loss.  Tracking in the PHENIX muon arms can
suffer in the presence of many tracks.  We embedded hits in the muon arms from
simulated single muons into real events and attempted to reconstruct the
simulated muons.  The efficiency for reconstructing muons in 0-20\% centrality
events was found to be approximately 20\% lower than the efficiency in 40-94\%
centrality events.  Additional particles were also included in the background
subtraction.  In addition to background from muons from hadron decays, hadrons
which make it through all of the absorber layers also must be subtracted, as
described in \cite{muon}.  Compared to the previous $p$+$p$ analysis, this
analysis also used a more careful calculation of the expected hadronic
background.  Particles which make it through only some of the absorber layers
are considered to be hadrons.  Simulations of hadronic interactions in the muon
arms are compared to the stopped hadron distributions, and the subtracted
background is tuned accordingly.  Regions of the detector in which the
simulation does not match the data well are not used in the analysis.

\begin{figure}[!h]
\centering
\includegraphics[width=0.45\textwidth]{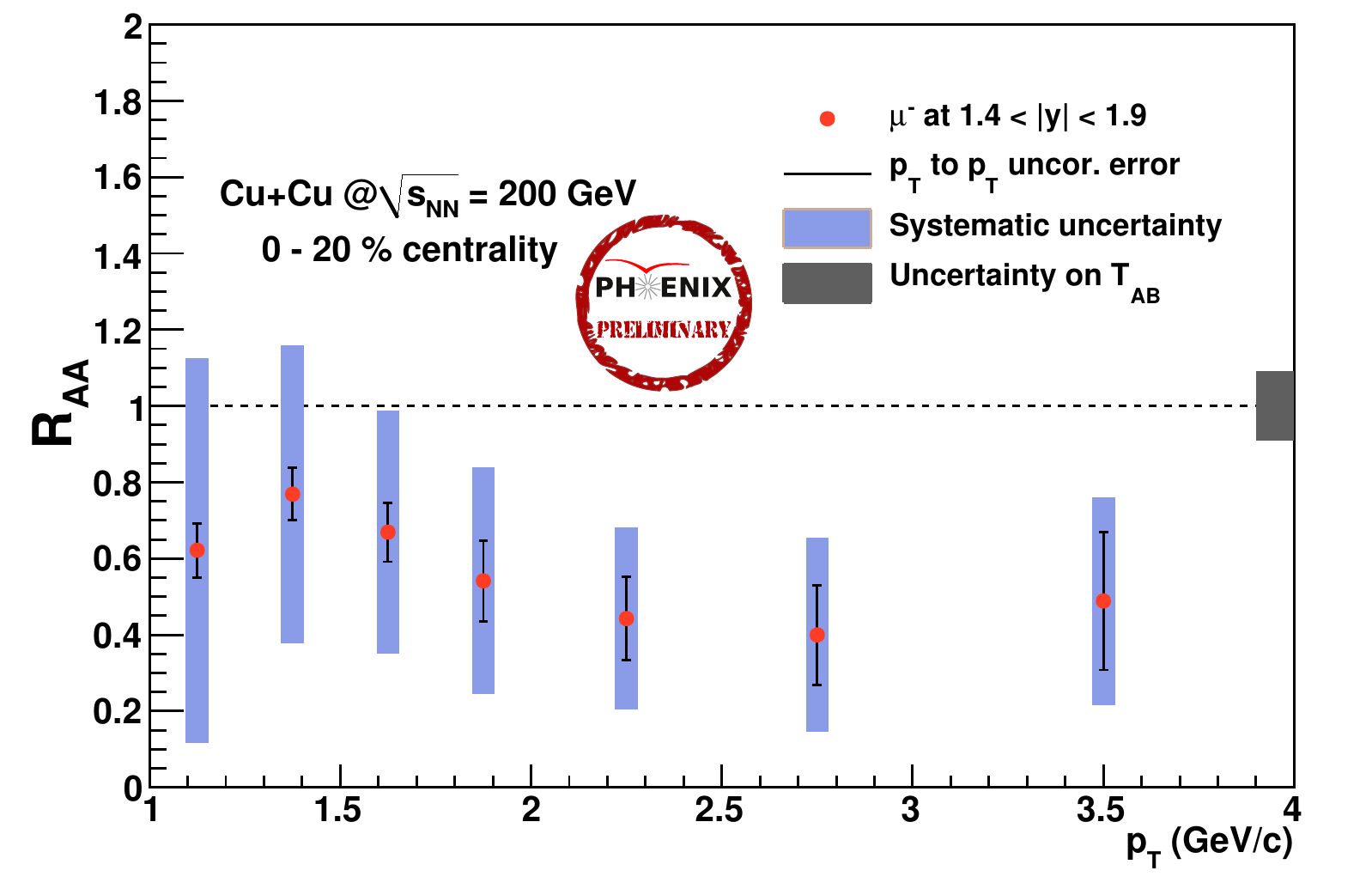}
\includegraphics[angle=90, width=0.45\textwidth]{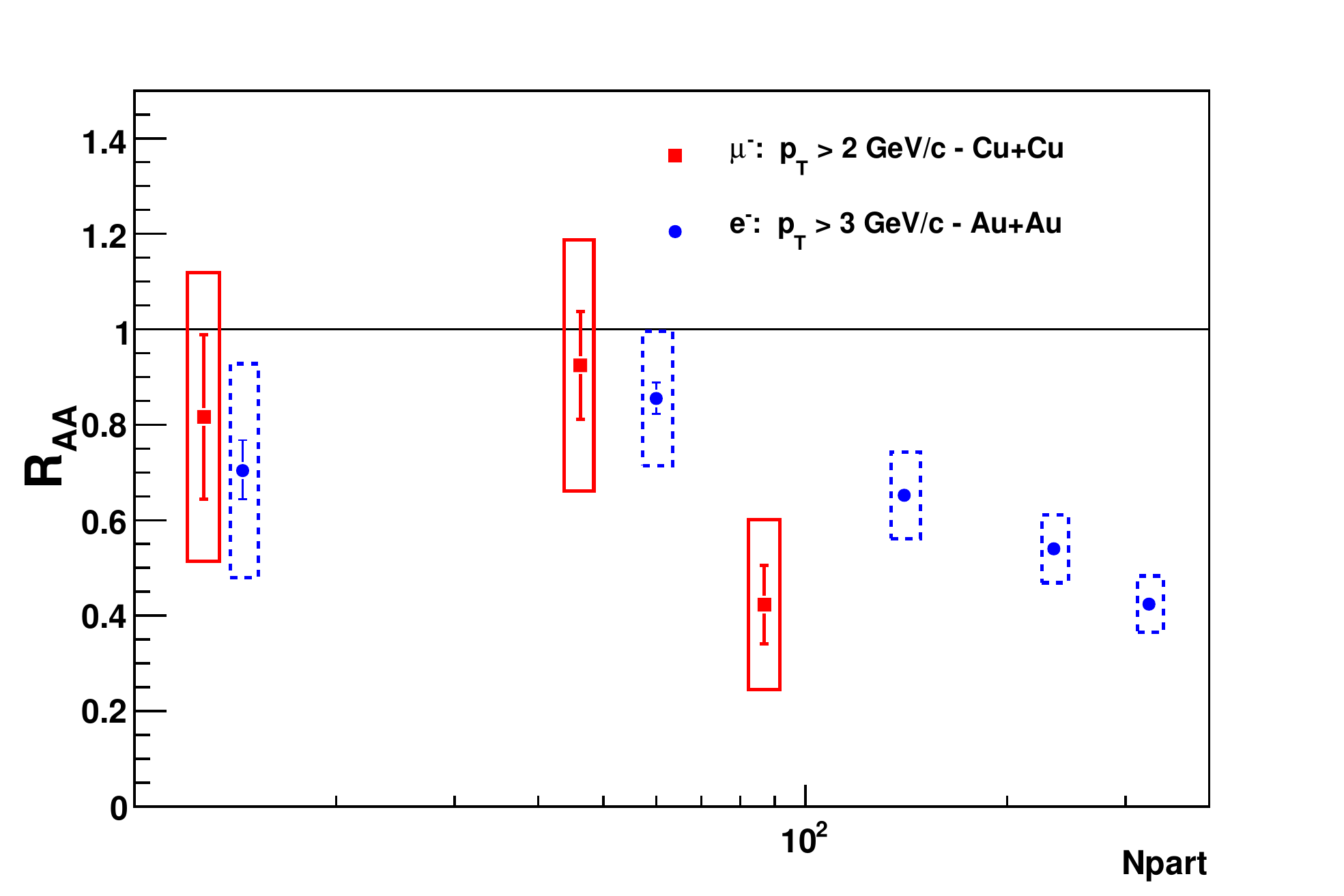}

\caption{left: $R_{AA}$ of prompt single muons in Cu+Cu collisions.  right:
$R_{AA}$ electrons from heavy flavor decays at midrapidity and of muons at
forward rapidity as a function of the number of participating nucleons in a
collision.}
\label{fig:mraa}
\end{figure}
Figure \ref{fig:mraa} shows the resulting $R_{AA}$ of prompt single muons in Cu+Cu collisions, using the $p$+$p$ measurement from \cite{donny} as a baseline.  Although the systematic errors are large, significant suppression of heavy flavor at forward rapidity can be seen in the 0-20\% centrality range.  It is interesting to note that the central value of the 0-20\% $R_{AA}$ for $p_T > $ 2 GeV/$c$ is smaller than that of electrons at mid-rapidity at the same $\textrm{N}_{\textrm{part}}$, which can be seen from Figure \ref{fig:mraa}.  This would be quite surprising if true, but we must await future detector and luminosity upgrades before being able to make a firm statement on this issue.





\begin{thebibliography}{00} 
   \bibitem{e_pp} A. Adare et al., {\it Phys. Rev. Lett.} {\bf 97}, 252002 (2006).

   \bibitem{e_aa} A. Adare et al., {\it Phys. Rev. Lett.} {\bf 98}, 172301 (2007).
   
   \bibitem{cesar} Cesar Luiz da Silva, fot the PHENIX collaboration, these proceedings.

   \bibitem{mt} A. Adare et al., {\it Phys. Lett. {\bf B}} {\bf 670}, 313 (2007).
   
   \bibitem{fonll} M. Cacciari, et al., {\it Phys. Rev. Lett.} {\bf 95}, 122001 (2005).
   
   \bibitem{jraa} A. Adare et al., {\it Phys. Rev. Lett.} {\bf 98}, 232301 (2007).
   
   \bibitem{donny} D. Hornback for the PHENIX collaboration, {\it J. Phys. G} {\bf 35}, 104113 (2008).
   
   \bibitem{muon} S.S. Adler et al., {\it Phys. Rev. {\bf D}} {\bf 76}, 092002 (2007).
   
   
   

\end{thebibliography}
\end{document}